\documentstyle[11pt,newpasp,twoside,epsf]{article}
\reversemarginpar

\begin{document}

\def\mic{$\mu$m\ }
\def\mj{M$_{\rm J}$\,}
\def\rj{R$_{\rm J}$\,}

\title{Models of Irradiated Extrasolar Giant Planets}
\author{Adam Burrows}
\affil{Department of Astronomy, The University of Arizona, Tucson, AZ, USA 85721}
\author{David Sudarsky}
\affil{Department of Astronomy, The University of Arizona, Tucson, AZ, USA 85721}

\begin{abstract}

We review some of the characteristics of irradiated extrasolar giant planets (EGPs),
in anticipation of their direct detection from the ground and from space.  Spectral measurements
are the key to unlocking their structural and atmospheric characteristics and to determining the true
differences between giant planets and brown dwarfs.  
In this spirit, the theoretical spectral and atmospheric calculations we 
summarize here are in support of the many searches for EGPs to
be conducted in the coming decade by astronomers
from around the world.

\end{abstract}

\section{Introduction}
\label{intro}

Since the discovery of 51 Pegasi b (Mayor and Queloz 1995) and the nearly
one hundred extrasolar giant planets (EGPs) that have been detected subsequently
by radial velocity techniques (see this proceedings and references in Burrows et al. 2001),
an increasing fraction of the world's astronomers has been engaged in determining
the best means to detect such planets directly.  While the orbital elements
of substellar-mass objects with M$_{\rm p}$sin(i)s that range from $\sim$0.2 to $\sim$10 \mj   
can constrain formation mechanisms 
and dynamical evolution, they are no substitute for direct spectral measurements.
It is by imaging the planet and obtaining optical, near-infrared, and mid-infrared
spectra that an EGP's atmospheric structure, composition, gravity, radius, and mass
can be determined.  Such physical characteristics are essential data if the study
of EGPs is to mature in the next decade into a major astronomical field.  They are
also essential if the distinctions between brown dwarfs and giant planets of the same
mass are to be determined.  It may be that, for a given primary 
star, different origins and histories at birth
translate into different compositions and rotation rates.  Spectra will be 
essential in determining this.

From space, SIM will provide accurate astrometric masses (not merely M$_{\rm p}$sin(i)s) for the
known EGPs.  However, from the ground spectral deconvolution techniques, adaptive 
optics, interferometry (e.g., using the LBT, VLTI, KIA)
and a host of promising and novel methods summarized during 
this conference encourage one to believe that light from
an EGP will soon be detected.  From space, optical 
coronagraphs with ultrasmooth mirrors (e.g., Eclipse, ESPI, JPF), infrared interferometers,
and precision transit missions (e.g., MOST, Eddington, Kepler, COROT) are in various stages of planning
or preparation.   The space-based transit missions will be preceded by a host
of ground-based transit searches (e.g., STARE, BEST, WASP, STEPSS, TeMPEST).
Transit data married with precise stellar and Doppler wobble measurements
can provide mass-radius relations for the close-in EGPs (``roasters")  
from which one can extract structural and evolutionary information (Guillot et al. 1996; Burrows et al. 2000).   
The first discovered transiting extrasolar planet, HD209458b,
for which a large radius of $\sim$1.4 \rj and a mass of 0.69 \mj 
were obtained (Cody and Sasselov 2002; Brown et al. 2001; Charbonneau et al. 2000;
Henry et al. 2000) has jump-started the scramble to understand 
EGPs under severe irradiation regimes.  The demonstration that the 
depth of HD209458b's transit is wavelength-dependent (Hubbard 
et al. 2001) and that neutral sodium resides in
its atmosphere (Charbonneau et al. 2002) is an indication of the vast potential of transit studies.
The discovery of a collection of transiting planets, not just one,  
will be a milestone in the study of EGPs.

This tempo of activity focussed on obtaining direct measurements of EGP properties 
demands a corresponding effort by the theoretical community to calculate the
spectra of EGPs around a variety of stars, at a variety of orbital distances,
and with a variety of masses and ages.  We have undertaken such a project and 
in this short contribution present some of our preliminary results.  A more comprehensive
treatment can be found in Sudarsky, Burrows, and Hubeny (2002, in preparation).
We have calculated EGP spectra for 51 Peg b, $\tau$ Boo b, HD209458 b, $\upsilon$ And b,c,d,
GJ 876 b,c, $\epsilon$ Eridani b, 55 Cnc b,c,d, HD114762 b, HD1237 b, and a host of
other radial-velocity EGPs, as well as for theoretical objects at the full potential
range of orbital distances, around a collection of stellar subtypes, and employing 
a variety of cloud models.  The fluxes at the Earth, as well as the planet/star contrasts
as a function of wavelength from 0.4 \mic to 30 \mic have been determined.

\section{A Potpourri of Irradiated EGP Spectra}
\label{pot}

\begin{figure}
\plotfiddle{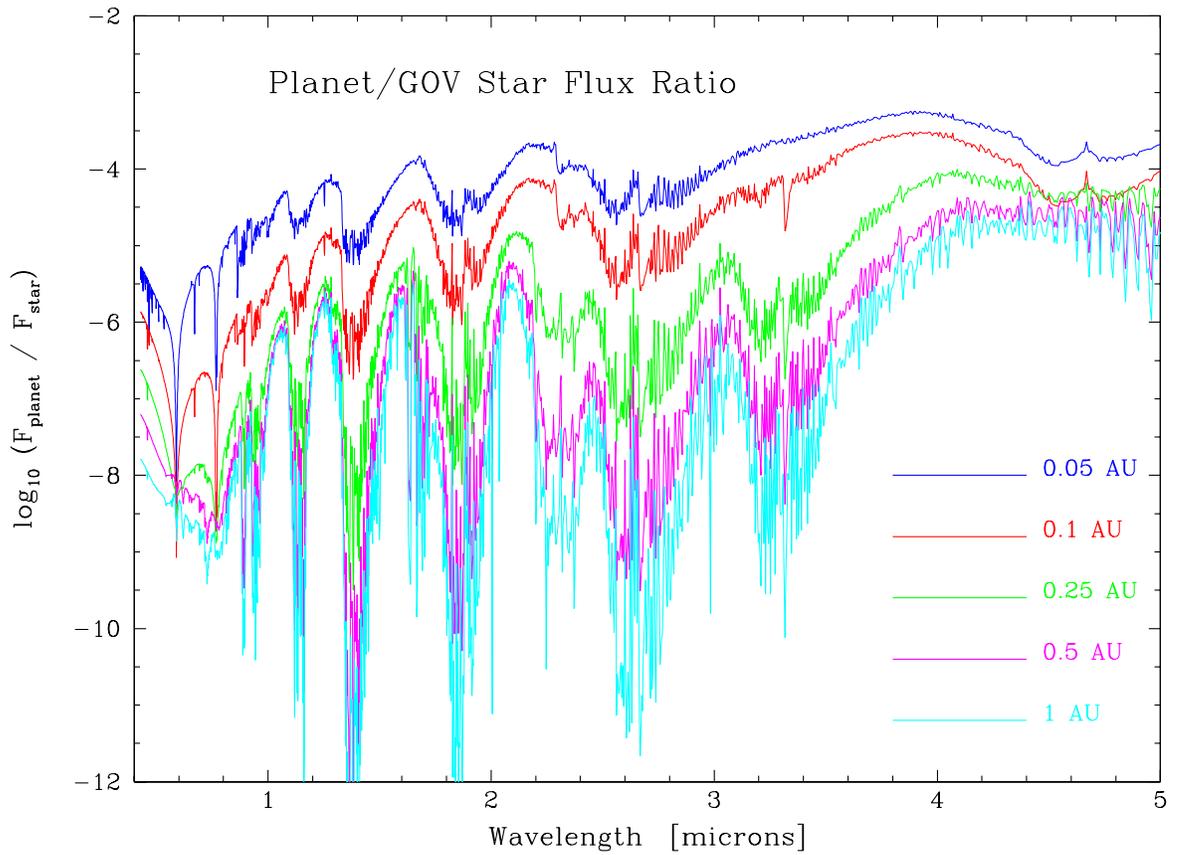}{6.0in}{-90}{60.0}{60.0}{-200}{400}
\caption{Theoretical planet/star flux ratios from 0.4 \mic to 5.0 \mic for
EGPs placed at orbital distances from 0.05 A.U. to 1.0 A.U.  These models
do not include the possible effects of clouds.}
\end{figure}

Figure 1 depicts the run with orbital distance from 0.05 to 1.0 A.U. of self-consistent EGP reflection spectra 
(including heating by stellar irradiation) around a G0 V star.   What is actually shown in Fig. 1
is the planet/star flux ratio, a quantity that in some ways is more useful to potential observers.
Rayleigh scattering at short wavelengths, methane features for the more distant (hence, cooler) objects, Na-D
and K I features at 0.589 \mic and 0.77 \mic, respectively, and water (steam) features are in evidence.  Without
Rayleigh scattering, the flux shortward of 0.6 \mic would be 3-6 orders of magnitude lower.  This set of
models does not include clouds (Sudarsky, Burrows, and Hubeny 2002), but the characteristic increase in the trough/peak
contrasts with increasing distance, the relative strength of the planet at 3-5 \mic, and the wide range of flux ratios
from $\sim 10^{-3}$ to $\sim 10^{-10}$ serve to emphasize that the choice of spectral range is crucial
if the direct light of the planet is to be measured.  In addition, 
though for roasters the planet flux at 10 parsecs
in the near infrared can reach $\sim$10 milliJanskys (a rather large number), 
the glare of the star is still a challenge to direct detection.   
Figure 2 portrays the corresponding temperature/pressure
profiles.  A signature of irradiation, the bump near one bar pressure, is in evidence.

\begin{figure}
\plotfiddle{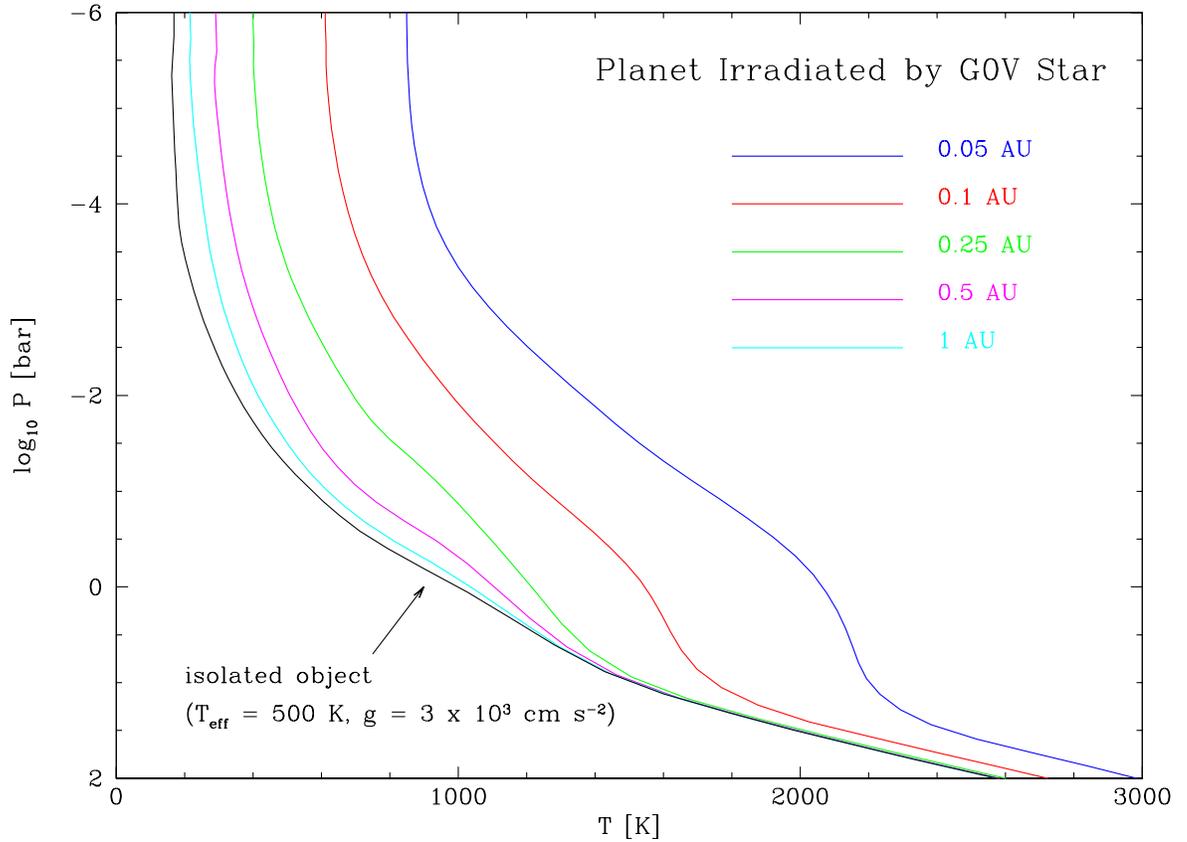}{6.0in}{-90}{60.0}{60.0}{-200}{400}
\caption{Profiles of the temperature (in Kelvin) versus pressure (in bars) for the models depicted in Fig. 1.
The inner boundary condition for this model set is an effective temperature
of 500 K.   A gravity of $3\times 10^{3}$ cm s$^{-2}$ was assumed.
Deep in the interior the models are convective.  The bump in the middle of the
profiles near one bar pressure reflects the effect of irradiation (heating) from outside.
}
\end{figure}

Figure 3 shows a representative theoretical spectrum at an average phase for
a planet that may be orbiting $\epsilon$ Eridani at 3.3 A.U. 
(Hatzes et al. 2000).  This model spectrum incorporates water clouds with 5-micron ice particles. 
As can be seen in Fig. 3,  the presence of clouds
elevates the flux in the optical to a significant degree.  The result is the classic ``two--hump"
spectrum (crudely, reflection at short wavelengths and emission at long wavelengths) (Saumon et al. 1996),
but with important differences.  In particular, there is a pronounced excess near 4--6 \mic
and the spectrum is peppered with CH$_4$ and water absorption features.

\begin{figure}
\plotone{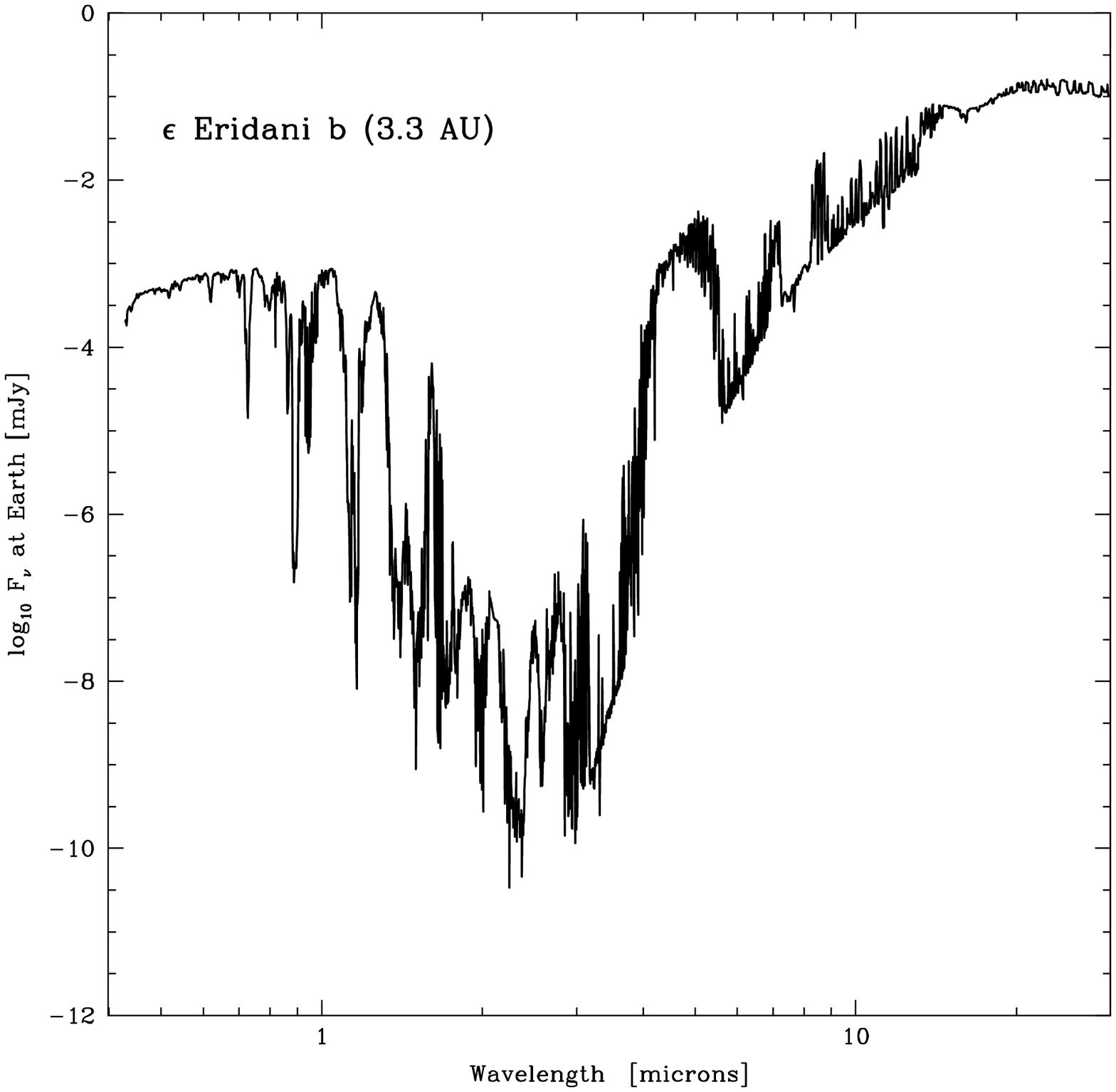}
\caption{A theoretical spectrum at average phase of the putative planet
orbiting $\epsilon$ Eridani at 3.3 A.U.  The star-planet angular
separation for this system is $\sim$1$^{\prime\prime}$.  A water cloud model 
with 5-micron ice particles is incorporated into
this calculation.  Note the presence of the pronounced feature around a wavelength of 5.0 microns, at which the
planet's spectrum may be near 10 microJanskys (as seen at Earth).
As suggested by the figure, the fluxes at longer wavelengths in the mid-IR
are expected to be even higher.
}
\end{figure}

For the close-in EGPs such as 51 Peg b, HD209458b, and $\tau$ Boo b,
the planet/star flux ratios shortward of one micron are $\sim$10$^{-5}$, but they climb to $3\times 10^{-4}$
around 1.65 \mic and 2.2 \mic.  At $\sim$4.0 \mic, they can be near $10^{-3}$, with a slight dependence 
on the presence of clouds.  The $\upsilon$ And system, boasting three EGPs with orbital distances from 0.059 A.U.
to 2.5 A.U., shows the entire range of potential behaviors for irradiated giant planets.
The corresponding flux ratios can vary from three to twelve orders of magnitude, depending
upon wavelength and cloud model.  The planets around GJ 876 (M4 V) provide an example
of irradiation by a dim star.  A transit by an EGP of such a late star (with a small radius)
would be spectacular in a way that the few nanoJansky fluxes in the near infrared from the wider
of the pair of GJ 876's planets is not.

\section{On the Wavelength Dependence of EGP Transits}
\label{wave}

HD209458b, while not a Rosetta Stone for the subject of EGPs, does nevertheless show 
the presence of atmospheric sodium and has provided a mass-radius point. More such points are anticipated.
The HST/STIS data for HD209458b (Brown et al. 2001; Charbonneau et al. 2002) cover the
wavelength range from 0.58 \mic to 0.64 \mic, which includes the Na-D line at 0.589 \mic.  To $\sim 4\sigma$,
a difference between the transit depth at Na-D and the average depth was detected.  However,
Na-D is not the only feature predicted in the transit depth spectrum.  Figure 4 from Hubbard et al. (2001)
shows that measurements of the transit depth as a function of wavelength can be used to reveal 
other atmospheric constituents such as potassium and, importantly, water.
``Absorption'' features appear upside-down on Fig. 4, where a large radius
reflects a larger opacity.  Therefore, a transit spectrum can be used as an erzatz
for a direct reflection spectrum (that may be more difficult to obtain) to determine atmospheric abundances.

\begin{figure}
\plotone{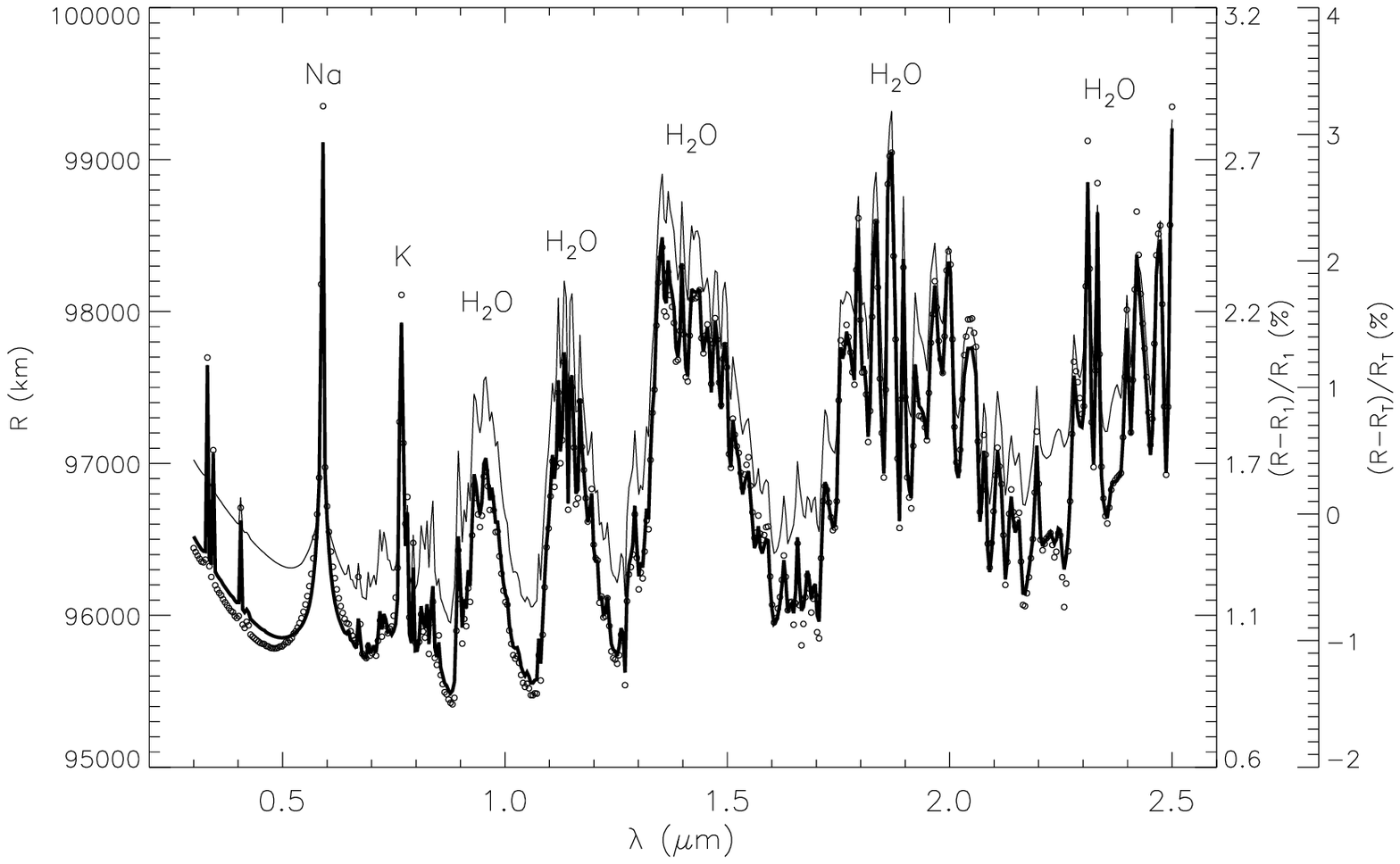}
\caption{Predicted variation of transit radius $R$
with wavelength ({\it heavy curve}, nominal $P-T$ profile;
{\it light curve}, cold $P-T$ profile; {\it open dots},
hot $P-T$ profile). The right-hand scales show, in percent,
the variation of $R$ with respect to $R_1$ and with
respect to $R_{\rm T}$, an $\sim$``average'' transit radius in
the visual wavelength band, adopted as
$R_{\rm T}=96500$ km $=1.35 R_{\rm J}$.  At wavelengths
where slant optical depth is high, $R$ is larger.
``Absorption'' features thus appear upside-down on
this plot.  Prominent features are labeled with the
responsible molecule (Figure taken from Hubbard et al. 2001).}
\end{figure}

However, care must be exercised in the interpretation of such ``spectra"; they are
not the same as reflection spectra.  In particular, the transit depth spectrum
is a probe of the slant column near the terminator of the planet, the cord optical depth.  
The actual spectrum of the planet is related more to the 
normal column depth and to the radial temperature/pressure
profile; it is an integral measure of whole-planet emissions.  
Fortney et al (2002, in preparation and presented at this conference)
interpret the discrepancy between the STIS Na-D results and the 
Hubbard et al. (2002) predictions at 0.589 \mic as being due to the neglect by the latter of photoionization
and charge exchange in HD209458b's upper atmosphere (down to $\sim$1--5 millibars).  
When Fortney et al. (2002) include the ionization of sodium in their transit spectrum calculation (which
involves 2D Monte Carlo transport), they can reproduce the transit observations.  But when we    
calculate the reflection spectra for the same object with and without ionization down
to the level necessary to reproduce the HST/STIS results, we see very little difference.    
In short, a transiting planet's reflection spectrum does not reveal much about its 
transit-depth spectrum and care must be taken not to confuse the two.

\section{Summary Thoughts}
\label{sum}

NASA and ESA are poised to spend 100's of millions of dollars in the next decade
to detect and characterize extrasolar planets.  The best data in the short term
will be for EGPs, not terrestrial planets.  Hence, for those most interested in
discovering Earths, the natural and inevitable path is by way of the 
giant planets now being discovered in profusion in the solar neighborhood.
For those for whom EGPs are not mere stepping stones, the next decade promises
a rich harvest of new worlds and stimulating finds.  Our calculations are designed to provide the 
necessary theoretical underpinnings for this quest at one of astronomy's newest frontiers.

\acknowledgments

The authors thank Bill Hubbard, Jonathan Lunine, 
Jim Liebert, Ivan Hubeny, Christopher Sharp, Drew Milsom, 
Maxim Volobuyev, Curtis Cooper, and Jonathan Fortney 
for fruitful conversations and help during the 
course of the work summarized here, as well as  
NASA for its financial support via grants NAG5-10760
and NAG5-10629.  They would also like to congratulate 
the organizers, Drake Deming and Sara Seager,
for pulling off without a perceptible hitch such a 
useful and important meeting.

\end{document}